\documentclass[11pt]{article}
\usepackage{epsf}
\newlength{\bredde}
\def\slash#1{\settowidth{\bredde}{$#1$}\ifmmode\,\raisebox{.15ex}{/}
\hspace*{-\bredde} #1\else$\,\raisebox{.15ex}{/}\hspace*{-\bredde} #1$\fi}
\newcommand{\beq}{\begin{equation}}
\newcommand{\eeq}{\end{equation}}
\newcommand{\noi}{\vspace{12pt}\noindent}
\newcommand{\lG}{\raise.3ex\hbox{$\stackrel{\leftarrow}{G}$}}
\newcommand{\lU}{\raise.3ex\hbox{$\stackrel{\leftarrow}{U}$}}
\newcommand{\lP}{\raise.3ex\hbox{$\stackrel{\leftarrow}{{\cal P}}$}}
\newcommand{\leta}{\raise.3ex\hbox{$\stackrel{\leftarrow}{\eta}$}}
\newcommand{\lOmega}{\raise.3ex\hbox{$\stackrel{\leftarrow}{\Omega}$}}
\newcommand{\ldr}{\raise.3ex\hbox{$\stackrel{\leftarrow}{\delta^r}$}}

\def\beqn{\begin{eqnarray}}
\def\eeqn{\end{eqnarray}}

\def\gtwid{\raise.3ex\hbox{$>$\kern-.75em\lower1ex\hbox{$\sim$}}}
\def\ltwid{\raise.3ex\hbox{$<$\kern-.75em\lower1ex\hbox{$\sim$}}}

\newcommand{\nr}[1]{(\ref{#1})}

\begin{document}
\topmargin -1.4cm
\oddsidemargin -0.8cm
\evensidemargin -0.8cm
\title{\Large{{\bf Staggered Fermions and Gauge Field Topology }}}

\vspace{1.5cm}

\author{~\\~\\
{\sc P.H. Damgaard$^a$, U.M. Heller$^b$, R. Niclasen$^a$} and 
{\sc K. Rummukainen$^{c,d}$}\\~\\~\\
$^a$The Niels Bohr Institute and $^c$NORDITA\\ Blegdamsvej 17\\ 
DK-2100 Copenhagen, Denmark\\~\\~\\
$^b$SCRI\\Florida State University\\Tallahassee, FL 32306-4130, USA\\~\\~\\
$^d$Helsinki Institute of Physics,\\
P.O.Box 9, 00014 University of Helsinki, Finland\\}
\maketitle
\vfill
\begin{abstract} 
Based on a large number of smearing steps, we classify SU(3) gauge field
configurations in different topological sectors. For each sector we
compare the exact analytical predictions for the microscopic Dirac operator
spectrum of quenched staggered fermions. In all sectors we find perfect
agreement with the predictions for the sector of topological charge zero,
showing explicitly that the smallest Dirac operator eigenvalues of
staggered fermions at presently realistic lattice couplings are insensitive 
to gauge field topology. On the smeared configurations, $4\nu$ eigenvalues
clearly separate out from the rest on configurations of topological charge
$\nu$, and move towards zero in agreement with the index theorem. 
\end{abstract}
\vfill
\begin{flushleft}
NBI-HE-99-24 \\
FSU-SCRI-99-42\\
NORDITA-1999/42 HE\\
hep-lat/9907019
\end{flushleft}
\thispagestyle{empty}
\newpage

\setcounter{page}{1}
\section{Introduction}

\noindent
The study of the relation between staggered fermions and (lattice) gauge
field topology has a long history, beginning with the work of Smit and
Vink \cite{SmitVink}. In particular, it is well known that staggered
fermions at realistic values of the gauge field coupling $\beta$
do not show the proper relation between fermion zero modes and gauge field
topology as it is dictated by the index theorem in the continuum. This
established problem with staggered fermions has recently surfaced again,
as lattice gauge theory studies have begun to test in detail the exact
analytical predictions for the microscopic Dirac operator spectrum
of staggered fermions \cite{V0,BBetal,DHK,EHN}. The crucial point is that
the whole microscopic Dirac operator spectrum, and not just that pertaining
to the exact zero modes, has been predicted to be very strongly dependent 
on the gauge field topology \cite{LS}. It falls into one of a set of 
universality classes that in addition depend on the gauge group, the number 
of fermion flavors and their color representation \cite{SV,ADMN,AD,OTV}.

\noi
In earlier studies \cite{V0,BBetal,DHK,EHN} of the spectrum of the smallest
staggered Dirac operator eigenvalues {\em all} gauge field configurations, 
irrespective of their possibly non-trivial winding numbers, were simply
bunched together, and the Dirac operator spectrum taken with respect
to this full average. This implies an implicit sum over all topological
sectors, for which the analytical predictions are very different from
those of sectors with fixed topological index \cite{top}. Nevertheless,
absolutely excellent agreement was found when comparing with just the sector 
of vanishing topological charge $\nu=0$. On the surface this may seem to
be just a simple consequence of the fact that also exact zero fermion modes
are missing in the staggered formulation. The issue is, however, much more
complicated. The lattice studies of refs. \cite{V0,BBetal,DHK,EHN} 
involved the distribution of just around 10 of the lowest Dirac operator
eigenvalues. Only positive eigenvalues were considered, since the
staggered Dirac spectrum has an exactly $\pm$ symmetric spectrum. Of these
few lowest eigenvalues, typically one would expect that up to 2-6
were actually the ``would-be'' zero modes, shifted away from the origin
by the staggered fermion artefacts. What should be the distribution of
these ``wrong'' small eigenvalues? In the Random Matrix Theory formulation
of the problem \cite{SV,ADMN} there is no answer to this question, as
there is no known way of imposing correctly almost-zero modes in the
theory. In that formulation one either {\em has} or {\em has not} exact zero
eigenvalues. This could incorrectly lead to the conclusion that as long
as staggered fermions do not produce {\em exact} zero modes, the distribution
of the smallest Dirac operator eigenvalues in that formulation will exactly
equal that of the $\nu=0$ sector. The argument is false, because as
$\beta$ is increased the appropriate number of the smallest Dirac operator
eigenvalues will slowly separate out, and move towards the origin. As this
happens, the whole microscopic Dirac operator spectrum will continuously
shift, and in this intermediate region there will certainly no longer be 
agreement with the analytical predictions for {\em any} sectors with 
$\nu \neq 0$, let alone the sum over all of them.

\noi
These considerations immediately raise the question of whether already at
the $\beta$-values considered in refs. \cite{V0,BBetal,DHK,EHN} there was
appreciable contamination of ``wrong'' non-zero eigenmodes. To settle
that issue, we report here on a high-statistics analysis (involving
around 17,000 gauge field configurations) of the microscopic Dirac operator
spectrum of SU(3) gauge theory with quenched staggered fermions. Correctly
classifying the gauge field configurations according to topology is
not a simple task, especially since by default we are excluded
from using fermionic methods. We have chosen to do the
classification according to the result of measuring the naive latticized
topological charge
\beq
\nu ~=~ \frac{1}{32\pi^2}\int\! d^4x {\mbox{\rm Tr}}[F_{\mu\nu}F_{\rho\sigma}]
\epsilon_{\mu\nu\rho\sigma} ~
\label{charge}
\eeq
on configurations obtained after a large number of so-called APE smearing
steps \cite{APE}, details of which will be given below. Once classified,
we have then measured the smallest Dirac operator eigenvalues on the
original {\em un-smeared} gauge field configurations. We in no way claim
that this is an optimal way in which to separate out the different topological
sectors, but it should at least have quite some overlap with other methods.
In particular, it is known, on average, to produce results very similar
to more conventional types of semi-classical cooling \cite{Colo}. 

\noi
Studying the fate of the smallest Dirac operator eigenvalues with staggered
fermions is here motivated by the need to understand the exact analytical
predictions for the microscopic Dirac operator spectrum. But the problem
is interesting also for other reasons. For instance, one needs to know
the extent to which staggered fermions correctly couple to gauge field
topology also if one wishes to compute, for instance, flavor singlet
pseudoscalar masses in lattice QCD. We believe that the microscopic Dirac
operator spectrum may be an excellent tool with which to assess how close
simulations with staggered fermions are to continuum physics related to
gauge field topology and the anomaly.

\noi
This short paper is organized as follows. In the next section we briefly 
describe how we implement the smearing technique on gauge links, and show
how this leads us to a classification of all gauge field configurations
into distinct sectors, labelled by the value of $\nu$. We also trace the
evolution of the smallest Dirac operator eigenvalues as a function of the
number of smearing steps, and show that these results are completely
in accord with expectations. In section 3 we compute the microscopic
Dirac operator spectrum in each of the different gauge field sectors,
and compare results to the analytical predictions. To zoom in on precisely
the {\em smallest} eigenvalue, which in topologically non-trivial sectors 
should behave very differently from the ordinary Dirac operator eigenvalues,
we also compare the distribution of just this smallest eigenvalue with the
exact analytical predictions. Finally, section 4 contains our conclusions.

\section{Analysis of Gauge Field Topology: APE-Smearing}

\noi
We perform the statistical analysis of the gauge field topology and
the Dirac operator eigenvalues using a total of 17454 SU(3) pure gauge
configurations with volume $V=8^4$ and lattice coupling
$\beta=6/g^2=5.1$.  We generate the configurations with an update
consisting of 4 microcanonical overrelaxation sweeps over the volume
followed by one pseudo--heat-bath update.  The configurations we analyze
are separated by 20 of these compound sweeps; this guarantees that the
configurations are effectively uncorrelated.

\noi
Let us give some technical details about the measurements of the
topological charge.  First, the naive topological density operator
\nr{charge}, when implemented on the lattice, is not truly
topological: it is sensitive to lattice ultraviolet modes, and, if
applied to the original lattice configurations, the (generally
non-integer) value of $\nu$ is completely dominated by the UV noise.
However, if the fields are smooth enough on the lattice scale, topology
can be uniquely defined.  

\noi
We make the original lattice fields smoother by applying repeated
APE-smearing:
during one smearing sweep, SU(3) link variables $U_\mu(x)$ are replaced by
\beq
   U_\mu(x) \rightarrow P_{\rm SU(3)} \left[ f U_\mu(x) + 
         \sum_{|\nu|\ne\mu} 
        U_\nu(x)U_\mu(x+\nu) U^\dagger_\nu(x+\mu) \right],
\label{eq:APE}
\eeq
where the sum over $\nu$ goes over both positive and negative
directions.  The operator $P_{\rm SU(3)}$ projects the $3\times3$
complex matrix $W_\mu(x)$, inside the brackets in (\ref{eq:APE}), to
SU(3), and $f$ is an adjustable parameter.  The projection is
performed by maximizing ${\rm Tr}\left(U^\prime_\mu(x)
W^\dagger_\mu(x)\right)$ over the SU(3) group elements
$U^\prime_\mu(x)$.  The smearing sweeps (consisting of a smearing of each 
link on the lattice) are performed up to 400 times.

\noi
While the APE-smearing is not particularly sensitive to the actual
choice of the parameter $f$, we used $f=7$ throughout our work.
This value is motivated by the comparison to the RG
blocking analysis by De Grand {\em et al} \cite{Colo}.  According to their
results, APE-smearing is quite effective in resolving instantons from
quantum fluctuations, while it preserves the long-distance properties
of the gauge field configurations much better than the standard cooling
algorithms.  Naturally, almost any cooling or smearing method will
destroy instantons with a size of order of the lattice spacing; however,
the lattice topology is not well defined at these length scales anyway.

\begin{figure}[t]
\centerline{\epsfysize=8cm\epsfbox{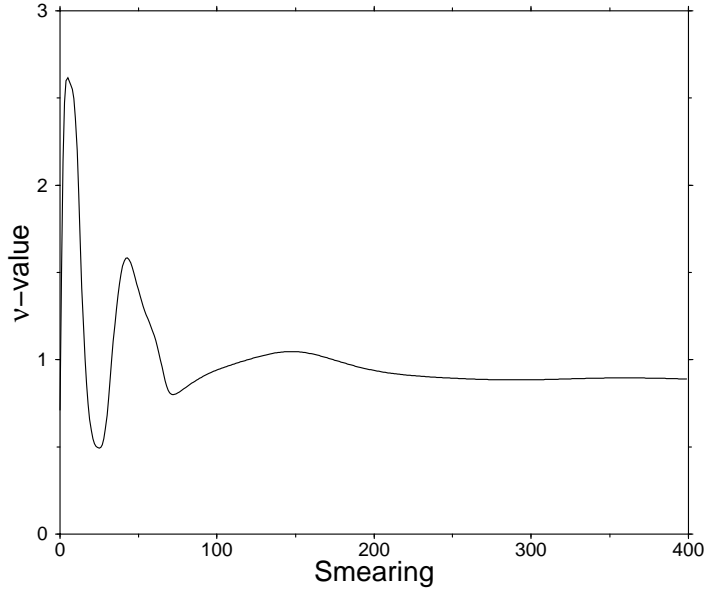}}
\vspace*{-2mm}
\caption[a]{
The topological charge of one configuration, measured
during the APE-smearing.  After initial `noise,' the charge measurement
quickly stabilizes to an almost-integer value.}
\label{fig:blocking}
\end{figure}

\noi
We calculate the field tensor $F_{\mu\nu}$ in Eq.~\nr{charge} at
lattice point $x$ by symmetrizing over the `clover' of the 4
($\mu,\nu$)-plane plaquettes which have one corner at point $x$.  In
Fig.~\ref{fig:blocking} we show a typical example of how the
topological charge (as measured with operator \nr{charge}) develops
during smearing.  As expected, the charge measurement is very noisy
when the configurations are rough, but after $\sim 100$ smearing
sweeps the measurement almost always stabilizes to almost an integer
value, which is then preserved for at least several hundreds of
smearing steps.

\begin{figure}[t]
\centerline{\epsfysize=8cm\epsfbox{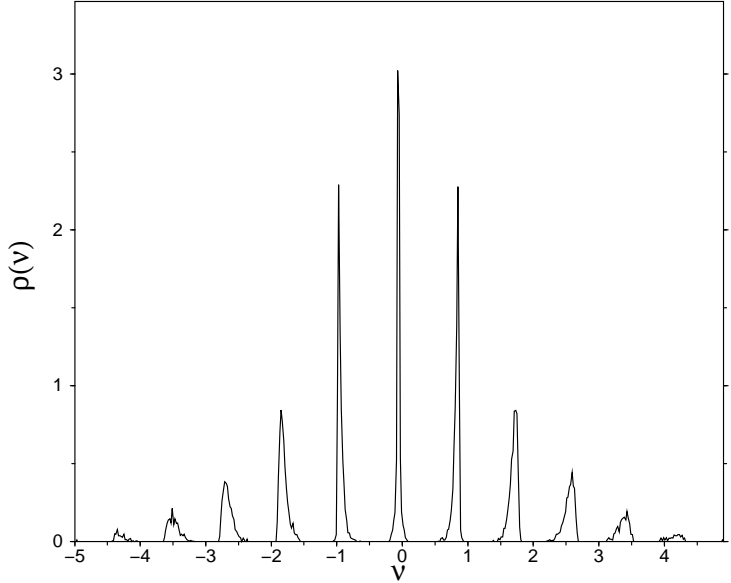}}
\vspace*{-2mm}
\caption[a]{
The distribution of the topological charge,
measured from 17454 $V=8^4$, $\beta = 5.1$ configurations.}
\label{fig:histogram}
\end{figure}

\noi
In Fig.~\ref{fig:histogram} we show the distribution of the
measured topological charge after 200 smearing steps.
The distribution is strongly peaked near integer values, as expected,
with a small downward drift of the peaks from exact integer values.
This is obviously caused by our naive $F\tilde F$ operator; using an
improved operator would presumably shift the peaks toward exact
integer values.  Nevertheless, this slight shift does not pose any
difficulties in assigning a topological index to the smeared
configurations. Indeed, it is fairly obvious from our figure that
there is simply a trivial ``renormalization'' of the naive topological
charge, which appears to be neatly quantized, not in units of
integers, but in units of around $\sim 0.8$.  If corrected with such a
prefactor of $\sim 1/0.8$, our distribution of topological charges
will fall on values very close to integers, and, as seen, the
distribution has the expected infinite-volume Gaussian shape.

\noi
We have measured the topological charge after 200 and
300 smearing sweeps, and accepted only configurations where the measured
charge remained in the neighborhood of the same ``renormalized'' integer 
value 0, $\pm 1$, $\pm 2$. In this way we have rejected configurations 
where the measured topological charge changed too much to be uniquely defined.
In this process we have rejected 37\% of our configurations, which simply 
have been discarded in all of the subsequent analysis.

\noi
Eigenvalues of the staggered Dirac operator
\beqn
\slash{D}_{x,y} &=& \frac{1}{2}\sum_{\mu}\eta_{\mu}(x)\left(U_{\mu}(x)
\delta_{x+\mu,y} - U^{\dagger}_{\mu}(y)\delta_{x,y+\mu}\right)\cr
&\equiv& \slash{D}_{e,o} + \slash{D}_{o,e}
\eeqn
are computed using the variational Ritz functional method \cite{Ritz}.
Here $\eta_{\mu}(x) = (-1)^{\sum_{\nu<\mu}x_{\nu}}$ are the staggered
phase factors. Letting $\epsilon(x) = (-1)^{\sum_{\nu}x_{\nu}}$, we have
explicitly indicated how $\slash{D}$ connects {\em even} sites, i.e. those
with $\epsilon(x)=+1$, with {\em odd} (those with $\epsilon(x)=-1$) ones, and
vice versa. We remind the reader that the staggered Dirac operator is
antihermitian with purely imaginary eigenvalues that come in pairs,
 $\pm i\lambda$. The operator $-\slash{D}^2$ is thus hermitian and
positive semi-definite, and the sign function $\epsilon(x)$ defined above
plays the role of $\gamma_5$ in the continuum ($i.e.$, this quantity
anticommutes with $\slash{D}: ~\{\slash{D},\epsilon\}=0$). Moreover, since
$-\slash{D}^2$ does not mix between even and odd lattice sites, it suffices
to compute the eigenvalues, on, say, the even sublattice. In fact, if
$\psi_e$ is a normalized eigenvector of $-\slash{D}^2$ with eigenvalue
$\lambda^2$, then $\psi_o \equiv \frac{1}{\lambda} \slash{D}_{o,e}\psi_e$
is a normalized eigenvector of $-\slash{D}^2$ with eigenvalue $\lambda^2$,
and non-zero only on odd sites. (Note that there is no difficulty with the
above definition of $\psi_o$, since we will never encounter
exact zero modes). In practice,
we make use of these properties, and compute only the (positive) eigenvalues of
$-\slash{D}^2$ restricted to the even sublattice, and then take the
(positive) square root. All eigenvalues 
to be shown in the following thus have an equal number of negative
companions, of the exact same magnitude.

\begin{figure}[t]
\centerline{\epsfysize=12cm\epsfbox{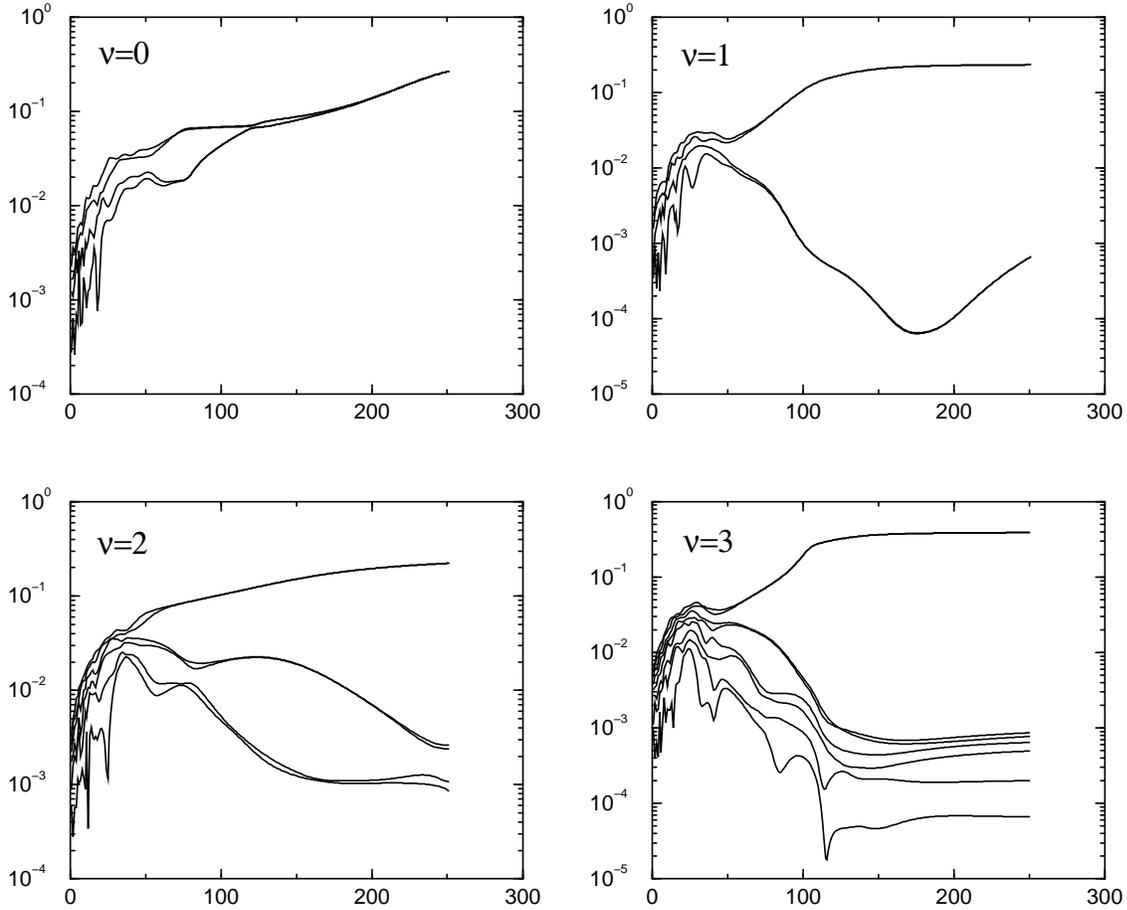}}
\vspace*{-2mm}
\caption[a]{
The behaviour of the lowest Dirac operator eigenvalues during the
APE smearing.  In each case there are $4\times\nu$ small eigenvalues; 
only the positive eigenvalues are shown here.}
\label{fig:blockvals}
\end{figure}

\noi
We have first investigated how the lowest eigenvalues behave 
during the smearing process described previously.  
As shown in Fig.~\ref{fig:blockvals}, the index theorem is seen
to be valid for smooth configurations: after $\sim 100$ smearing sweeps,
we have exactly $4\times \nu$ very small eigenvalues (corresponding to
4 continuum flavors for each staggered fermion flavor), whereas the other
eigenvalues grow larger.  This implies that in the continuum limit
the eigenvalues of the staggered Dirac operator should indeed 
depend on the topology of the gauge configuration. As expected, on the
smooth configurations the counting of flavors is precisely as in the
continuum limit: we observe 4 flavor degrees of freedom associated with
the staggered Dirac operator.

\section{The Microscopic Dirac Operator Spectrum}

\noi
Having described our procedure for classifying our gauge field configurations
according to their smeared topological charge $\nu$, we now proceed to
measure the distribution of the smallest Dirac operator eigenvalues. We
stress that we use the smearing technique {\em only} as a means of defining
the classification of the original un-smeared gauge field configurations.
All measurements presented below are performed on the original, 
{\em un-smeared}, gauge field configurations, after having discarded
those that failed to be classified according to the criterion described
above.

\noi
The spectral density of the Dirac operator is given by
\beq
\rho^{(\nu)}(\lambda) ~\equiv~ \langle \sum_n\delta(\lambda -
 \lambda_n)\rangle_{\nu}
\eeq
in each sector of topological charge $\nu$. The associated microscopic
Dirac operator spectrum is defined by enhancing the smallest eigenvalues
according to the size of the lattice space-time volume $V$. Let
\beq
\Sigma ~=~ \lim_{m \to 0}\lim_{V\to \infty} \langle\bar{\psi}\psi\rangle
\eeq
denote the infinite-volume chiral condensate as defined in the conventional
manner. One then blows up the small eigenvalues by keeping $\zeta \equiv
\lambda\Sigma V$ fixed as $V\to\infty$, and introduces the microscopic
spectral density \cite{SV}
\beq
\rho_s^{(\nu)}(\zeta) ~\equiv~ \frac{1}{V}\rho^{(\nu)}\left(
 \frac{\zeta}{\Sigma V} \right) ~,
\eeq
which then again is measured in each topological sector. For the case at
hand, the microscopic spectral density has been computed from both 
Random Matrix Theory \cite{SV} and from the effective finite-volume partition 
functions \cite{AD,OTV}. The results agree, and the simple analytical result 
for the quenched theory ($J_n(x)$ is the $n$th order Bessel function),
\beq
\rho_s^{(\nu)}(\zeta) ~=~ \pi\rho(0)\frac{\zeta}{2}\left[J_{\nu}(\zeta)^2
+ J_{\nu-1}(\zeta)J_{\nu+1}(\zeta)\right] 
\label{rhobessel}
\eeq
is exact in the limit where $V\to \infty$ and $V \ll 1/m_{\pi}^4$. Here
$\rho(0)$ is the macroscopic spectral density of the Dirac operator,
evaluated at the origin. By the well-known Banks-Casher relation, it
is related to the chiral condensate through $\pi\rho(0) = \Sigma$. 

\begin{figure}[t]
\centerline{\epsfysize=12cm\epsfbox{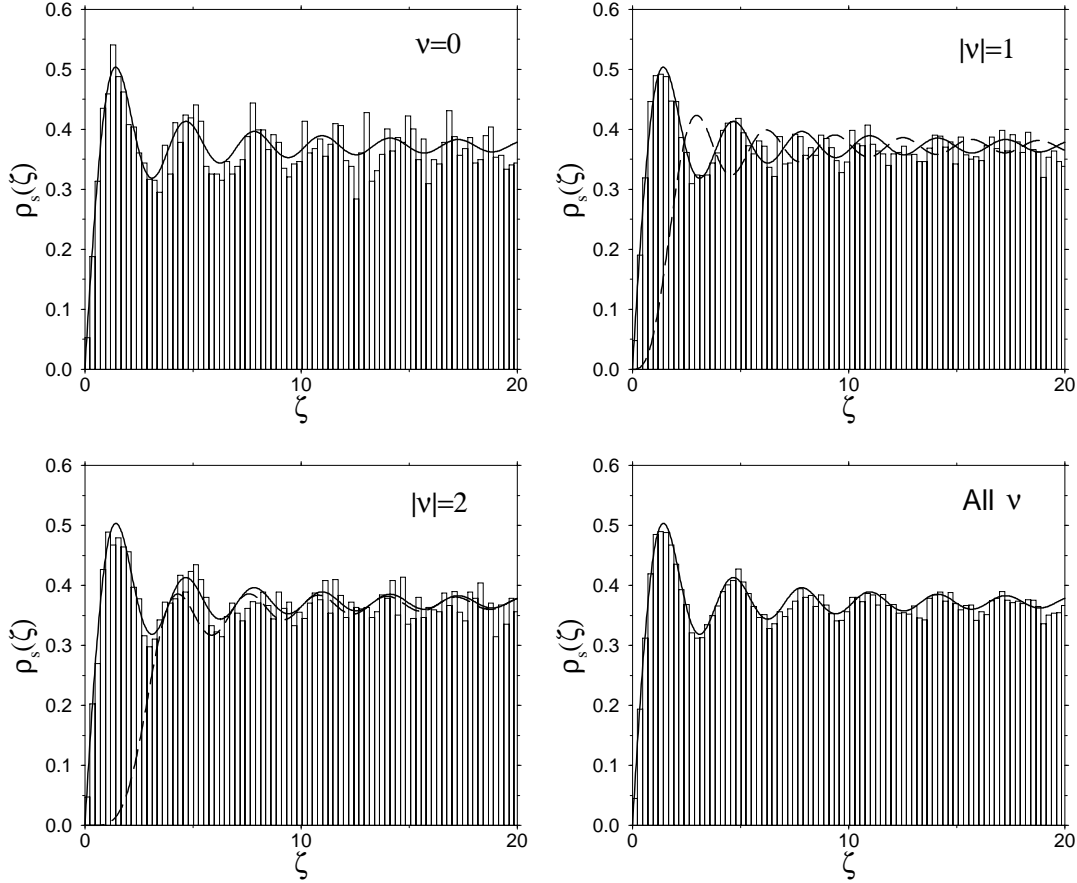}}
\vspace*{-2mm}
\caption[a]{
The microscopic spectral density of the Dirac operator for topological
sectors $\nu=0$, 1 and 2, and for all of the configurations combined.  
The continuous line is the theoretical prediction for the
sector $\nu=0$, and the dashed lines are the predictions for
$|\nu|=1$ and 2.}
\label{fig:rho}
\end{figure}

\noi
In Fig.~\ref{fig:rho} we compare these analytical predictions for the
sectors of $\nu = 0, 1$ and 2. We did not have enough statistics to
perform a similar analysis on what we would classify as $\nu=3$
configurations; however, as will be evident, there was no need to do
this either.  Our first comment concerns statistics. Because the
configurations with $+\nu$ and $-\nu$ should give rise to the same
microscopic Dirac operator spectrum, we actually end up with better
statistics for the $\nu=\pm 1$ configurations combined. In total, we
had 2683 configurations labelled as $\nu=0$, 4797 configurations
labelled as $\nu=\pm 1$, and 3493 configurations labelled as $\nu =
\pm 2$. Indeed, we observe from Fig.~\ref{fig:rho} the curious fact
that the agreement between the analytical curve for the microscopic
spectral density of the staggered Dirac operator actually seems to be
poorer on the $\nu=0$ than on the $\nu=\pm 1$ configurations.  We
attribute this solely to statistical fluctuations. It is quite obvious
that even configurations classified as of $\nu=\pm 1$ and $\nu=\pm 2$
topological charge give rise to a microscopic staggered Dirac operator
spectrum here which is indistinguishable from that of the $\nu=0$
configurations. The agreement with the exact analytical formula for
$\nu=0$ configurations (\ref{rhobessel}) is extraordinarily good on
all three classes of configurations. We have also indicated on the
figure the predictions for $|\nu|=1$ and $|\nu|=2$; there is clearly
no way our Monte Carlo data can be compatible with these predictions.
Combining all data, we obtain perfect agreement with the analytical $\nu
=0$ prediction, with very small statistical errors.

\noi
To focus more closely on just the smallest eigenvalue, we have also
compared its distribution with the analytical predictions for different
topological sectors \cite{NDW}. Let us denote the distribution of the
lowest (rescaled) eigenvalue in a sector of topological charge $\nu$ by 
$P^{(\nu)}(\zeta)$. From the general formula of ref. \cite{NDW} one
finds for the quenched theory, with our normalization convention:
\beqn
P^{(0)}(\zeta) &=& \pi\rho(0)\frac{\zeta}{2}e^{-\zeta^2/4} \cr
P^{(1)}(\zeta) &=& \pi\rho(0)\frac{\zeta}{2}I_2(\zeta)e^{-\zeta^2/4} \cr
P^{(2)}(\zeta) &=& \pi\rho(0)\frac{\zeta}{2}\left[I_2(\zeta)^2
- I_1(\zeta)I_3(\zeta)\right]e^{-\zeta^2/4} ~.
\eeqn

\begin{figure}[t]
\centerline{\epsfysize=12cm\epsfbox{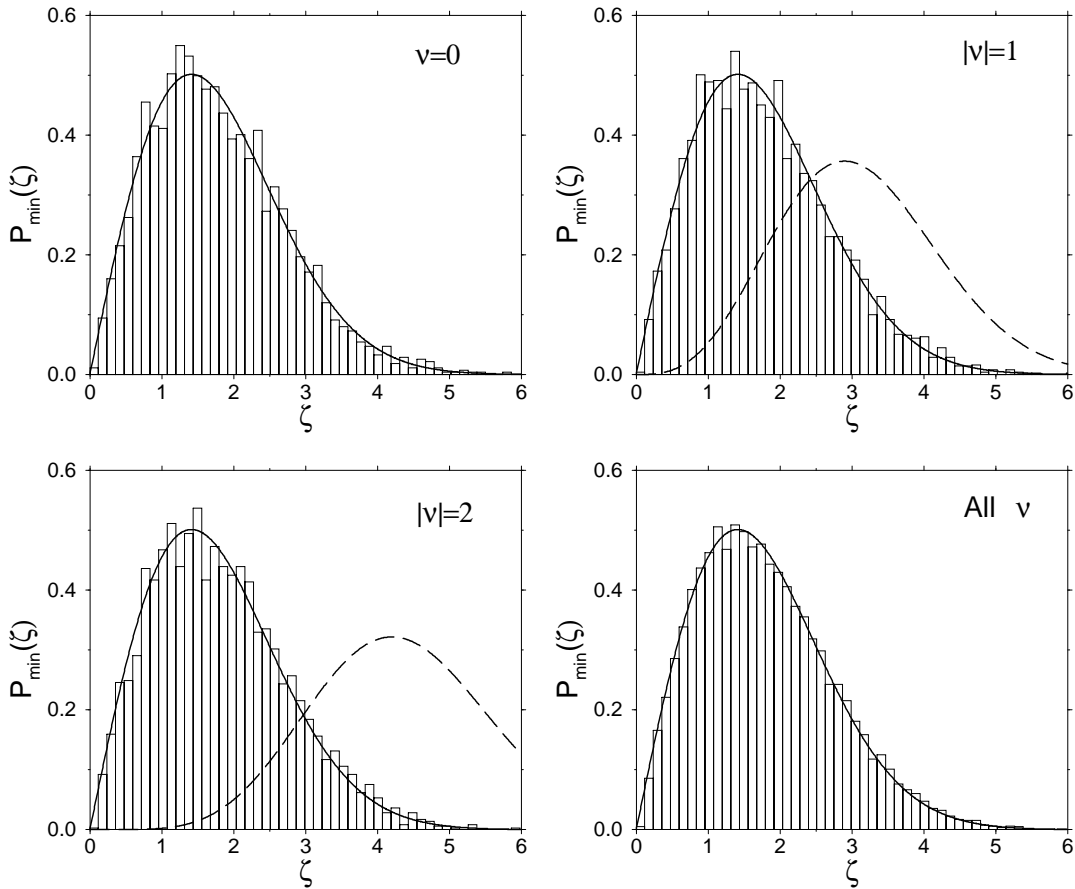}}
\vspace*{-2mm}
\caption[a]{
The distribution of the lowest eigenvalue of the Dirac operator for
topological sectors $\nu=0$, 1, 2, and for the sum over all
configurations.  Continuous (dashed) lines show the theoretical
prediction for $\nu=0$ ($|\nu|=1,2$).}
\label{fig:Pmin}
\end{figure}

\noi
Checking the distribution of just the smallest eigenvalue is obviously
the most sensitive test of whether there is any appreciable
contamination of would-be zero modes in the different topological
sectors. We show in Fig.~\ref{fig:Pmin} the lowest eigenvalue
distributions in the three different topological sectors, and compare
them with the analytical predictions.  Clearly no deviations 
are seen at all from the $\nu=0$ prediction, in all sectors.

\noi
Finally, one could well ask what would happen if we instead measured
the distribution of the smallest staggered Dirac operator eigenvalues
on the {\em smeared} configurations, after removing by hand those
eigenvalues that obviously should be classified as zero modes. All
indications are of course that such smeared configurations should produce
the correct behavior of the eigenvalues in the different topological
sectors. The reason why we have not performed such measurements on
the smoothened configurations is that the ensemble average is not a
very meaningful concept in that case. The results are clearly very sensitive
to where we cut the APE-smearing procedure, and even with a fixed number
of smearings it is not at all obvious that this could provide
us with a sensible ensemble average.

\section{Conclusions} 

\noi
At a lattice coupling of $\beta=5.1$ the microscopic Dirac operator spectrum
of staggered fermions in quenched SU(3) lattice gauge theory displays
{\em no deviations at all} from the analytical prediction of the $\nu=0$ 
topological sector. Even after roughly classifying all gauge field 
configurations into different topological sectors by means of the value
of $F\tilde{F}$ on sufficiently smeared configurations, we have found no
deviations from the $\nu=0$ predictions, in any of the different sectors.  
In view of earlier results, which simply bunched all gauge field
configurations together independently of their proper assignments of
topological charge \cite{V0,BBetal,DHK,EHN}, and which found excellent
agreement on this total sample of configurations, these results are
not extremely surprising. It is nevertheless very disturbing that
at the presently probed lattice couplings even gauge field configurations
that clearly should be classified as carrying non-trivial topological
charge $\nu$ are not at all seen as such by staggered fermions. The agreement
with the analytical predictions for the microscopic Dirac operator spectrum
in a gauge field sector of topological charge zero is thus a great
success of the analytical framework \cite{SV,ADMN,AD,OTV}, but it also
indicates a clear failure of the staggered fermion formulation at these
lattice couplings. What it means is that Monte Carlo simulations with 
staggered fermions at these lattice couplings are oblivious to net gauge field
topology, an essential ingredient of the dynamics of non-Abelian gauge 
theories. 
Such simulations are simply not mimicking the correct path integral of
the continuum theory. It is comforting that more sophisticated fermion
formulations that correctly build in the broken/unbroken chiral Ward
identities in sectors of fixed gauge field topology now are feasible 
alternatives. There are already clear results which show that these new
fermion formulations correctly reproduce the analytical predictions for the
microscopic Dirac operator spectrum in sectors of non-trivial gauge
field topology \cite{Lang,EHKN,DEHN}.

\noi
One surprising aspect of the present work is that it shows that the typically 
2-6 would-be zero modes of staggered fermions at these lattice spacings
behave entirely as ``conventional'' (small) Dirac operator eigenvalues
on topologically trivial configurations. In particular, although they consist
of up to 50\% of all the small eigenvalues we probe here, their distributions
fall entirely on top of the conventional small eigenvalues. Certainly,
these would-be zero modes will eventually, as the lattice spacing is
decreased, separate out, and completely distort the microscopic Dirac
operator spectrum. This deformation of the smallest eigenvalue spectrum
may already have been seen in the 2-d Schwinger model \cite{Hip}. It is
challenging to search for the corresponding onset of correct topological
properties with staggered fermions in this SU(3) gauge theory at smaller
lattice spacings. The microscopic Dirac operator spectrum is an excellent
tool with which to measure this in a precise and quantitative manner.

\vspace{1cm}

\noi
{\sc Acknowledgements:} The work of P.H.D. and K.R. has been partially 
supported by EU TMR grant no. ERBFMRXCT97-0122, and the work of U.M.H.
has been supported in part by DOE contracts DE-FG05-85ER250000 and
DE-FG05-96ER40979. In addition, P.H.D. and U.M.H. 
acknowledge the financial support of NATO Science Collaborative Research
Grant no. CRG 971487 and the hospitality of the Aspen Center for Physics.


\end{document}